\documentclass[lettersize,journal]{IEEEtran}
\usepackage{amsmath,amsfonts}
\usepackage{algorithmic}
\usepackage{algorithm}
\usepackage{array}
\usepackage[caption=false,font=normalsize,labelfont=sf,textfont=sf]{subfig}
\usepackage{textcomp}
\usepackage{stfloats}
\usepackage{url}
\usepackage{verbatim}
\usepackage{graphicx}
\usepackage{cite}
\usepackage{threeparttable}
\begin{document}
	
	\title{Passive Human Sensing Enhanced by Reconfigurable Intelligent Surface: Opportunities and Challenges}
	
	\author{Xinyu Li,~\IEEEmembership{Member,~IEEE}, Jian Wei You,~\IEEEmembership{Senior Member,~IEEE}, Ze Gu, Qian Ma,~\IEEEmembership{Member,~IEEE},\\ Long Chen, Jingyuan Zhang, Shi Jin,~\IEEEmembership{Senior Member,~IEEE}, and Tie Jun Cui,~\IEEEmembership{Fellow,~IEEE}
		\thanks{Xinyu Li, Jian Wei You, Ze Gu, Qian Ma, Long Chen, Jingyuan Zhang and Tie Jun Cui are with the State Key Laboratory of Millimeter Wave, Southeast University, Nanjing 210096, China (E-mail: \{xinyuli, guze, maqian, 220220703,zhangjingyuan, jvyou, tjcui\}@seu.edu.cn).}
		\thanks{Shi Jin is with the National Mobile Communications Research Laboratory, Southeast University, Nanjing 210096, China (E-mail: jinshi@seu.edu.cn). \\ \textit{(Corresponding author: Jian Wei You; Tie Jun Cui)}.}
		}
	
	
	
	\maketitle
	
	\begin{abstract}
		Reconfigurable intelligent surfaces (RISs) have flexible and exceptional performance in manipulating electromagnetic waves and customizing wireless channels. These capabilities enable them to provide a plethora of valuable activity-related information for promoting wireless human sensing. In this article, we present a comprehensive review of passive human sensing using radio frequency signals with the assistance of RISs. Specifically, we first introduce fundamental principles and physical platform of RISs. Subsequently, based on the specific applications, we categorize the state-of-the-art human sensing techniques into three types, including human imaging, localization, and activity recognition. Meanwhile, we would also investigate the benefits that RISs bring to these applications. Furthermore, we explore the application of RISs in human micro-motion sensing, and propose a vital signs monitoring system enhanced by RISs. Experimental results are presented to demonstrate the promising potential of RISs in sensing vital signs for manipulating individuals. Finally, we discuss the technical challenges and opportunities in this field. 
	\end{abstract}
	
	
	
	\section{Introduction}
	With the advancement of wireless technologies, passive human sensing based on wireless signals has achieved remarkable success and found applications in various areas, including health care, surveillance and ambient assisted living. Such a passive human sensing technique is realized by capturing the variations in reflected signals (e.g., radio absorption, scattering, and polarization) caused by the presence and movements of human targets \cite{li2022integrated}. To quantify the impact of human bodies on wireless channels for sensing purposes, off-the-shelf signal measurements, including the received signal strength indicator (RSSI) and channel state information (CSI), are commonly utilized \cite{liu2019wireless}. Due to the non-intrusive, non-cooperative and privacy-protecting superiorities, radio frequency (RF) based sensing has attract significant attention over the past decade. Moreover, with the widespread deployment of mobile communication infrastructure, RF-based human sensing could become more ubiquitous and precise, particularly in indoor environments where high coverage could be achieved.
	
	However, current wireless human sensing systems face certain fundamental challenges, limiting their applicability to the real world \cite{liu2019wireless}. For instance, multipath effects often lead to heavy attenuation of echo signals reflected by human bodies, making them sensitive to environmental noises. As a result, the sensing range and accuracy are compromised. Meanwhile, the non-line-of-sight (Non-LOS) human sensing is challenging and even unfeasible due to the absence of a direct wireless link between the transmitter (Tx) and the receiver (Rx). Furthermore, existing wireless sensing systems (e.g., WiFi and portable radar) have limited degrees of freedom and small aperture, resulting in low time/spatial resolution. This limitation leads to the mixing of reflected signals from different components at the receiver, further degrading the overall sensing performance. To fix these issues, we contemplate the following question: \textit{can we dynamically control signal propagation and elaborate customized radio channels for human sensing purposes?} 
	
	\begin{figure*}[h]
	\centering
	\includegraphics[width=1.5\columnwidth]{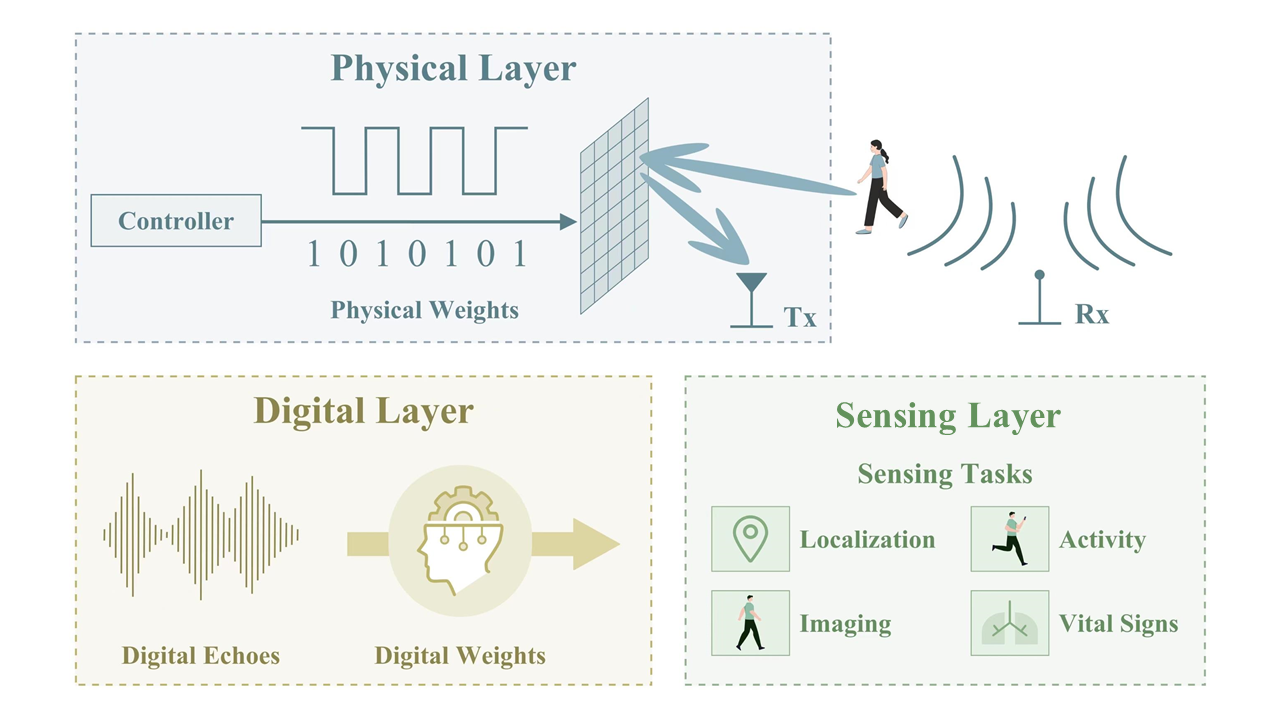}
	\caption{Schematic of RIS-enhanced passive human sensing, which is composed of the physical layer, the digital layer and the sensing layer.}
	\label{fig3}
    \end{figure*}

	The reconfigurable intelligent surface (RIS) \cite{9206044}, composed of a two-dimensional array of meta-atoms and renowned for its exceptional electromagnetic wavefront manipulation capabilities, holds great promise in addressing the aforementioned challenges. By individually programming the phase and amplitude of each reconfigurable meta-atom, the RIS can flexibly adjust unwanted propagation channels into desirable ones to facilitate high-quality passive human sensing. Compared with the conventional RF-based passive human sensing systems, the RIS-aided system incorporates a unique trainable physical layer composed of the reconfigurable metasurface, as shown in Fig. \ref{fig3}. Such an alteration enables customization of both the physical (the RIS) and the digital (the sensing algorithm) layers, thereby promoting the performance of passive human sensing at the physical and digital levels \cite{li2022intelligent}.
	
	In this paper, we present a comprehensive review of passive human sensing enhanced by RISs. We first provide a brief introduction to the background and implementation of RIS for RF-based human sensing. Subsequently, we place our main focus on three sensing tasks, i.e., imaging and segmentation, localization and tracking, and human posture recognition. Additionally, we provide an overview of existing RIS-aided passive human sensing systems. Furthermore, beyond the three typical tasks, some preliminary experimental results are provided to demonstrate the potential applications of RISs in multi-person micro-motion sensing. Finally, we discuss various challenges and opportunities for future investigations in this field before presenting concluding remarks.
	
	\section{Background and Implementation of RISs}

	\subsection{Background of RISs}
	RIS technology is enabled by metasurfaces, which are two-dimensional (2D) artificial surfaces composed of periodic or quasi-periodic meta-atoms in subwavelength scale \cite{9206044}.
	To date, the overwhelming majority of RISs used in RF-based human sensing have employed the space-domain-coding (SDC) pattern. For SDC RISs, the coding sequences are generally optimized in the spatial domain and fixed in time, switching with the command of the control system \cite{zhang2018space}. The advantages of employing SDC RISs for sensing can be summarized as follows. Firstly, the SDC RISs enable the customization of propagation channels to enhance sensing capabilities, and make Non-LOS sensing feasible. Secondly, SDC RISs aid in focusing sensing signals on specific areas of interest, significantly reducing interferences from irrelevant regions and improving signal-to-noise ratio (SNR) for long-range sensing. Additionally, SDC RIS is able to break through the resolution limitation of the existing wireless sensing systems with a simple hardware structure, empowering wireless systems with greater sensing capabilities. Thanks to these distinctive advantages, SDC RISs have been effectively applied to a wide range of human sensing tasks, leading to significant achievements in this field.
	
	Furthermore, by jointly encoding the parameters (e.g., reflection amplitudes and phases) of RISs in time and space, the EM waves could be manipulated in both the spatial and spectral domains \cite{zhang2018space}. By this means, one can simultaneously manipulate the harmonic distribution (frequency domain) and the propagation direction (spatial domain) of EM waves. Although spatial-time-coding (STC) RISs have made significant progress and found applications in various fields such as wireless communications, mobile user localization, and tracking, they are seldom utilized for passive human sensing tasks. Nevertheless, there is promising potential om employing STC RISs to overcome the current challenges in passive human sensing and achieve performance improvements, which will be introduced in Section \ref{Experiments}.

	\begin{figure}[h]
	\centering
	\includegraphics[width=1.1\columnwidth]{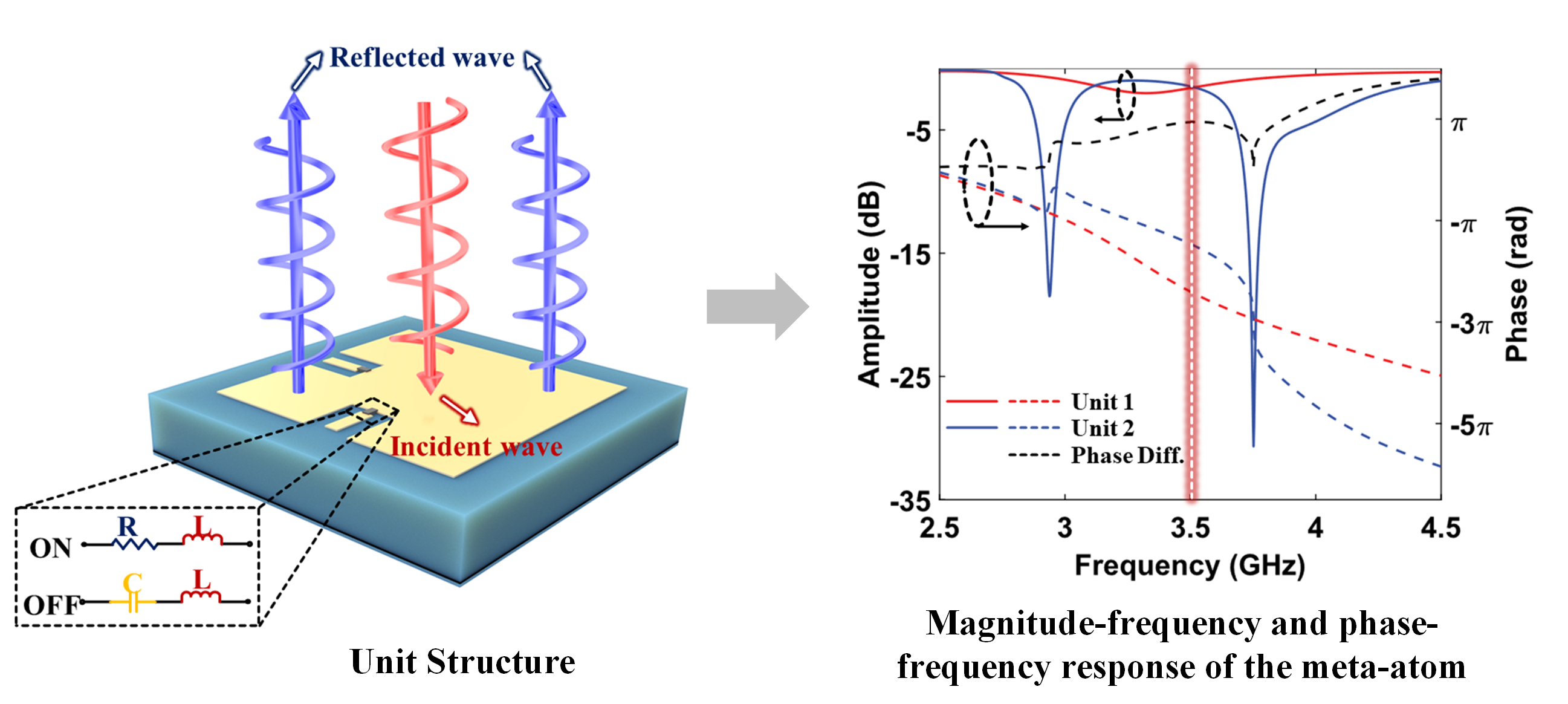}
	\caption{The 1-bit RIS implementation, and the corresponding magnitude-frequency and phase-frequency response of the designed meta-atom. }
	\label{fig6}
    \end{figure}

	\subsection{RIS Implementation}
	A 1-bit RIS operating at $\sim$3.5 GHz is designed as shown in Fig. \ref{fig6}. CST Microwave Studio is used to design the meta-atom. The digital meta-atom includes two PIN diodes, each of which is integrated to electrically and dynamically control the reflected EM response. Furthermore, from the magnitude-frequency and phase-frequency response of the designed meta-atom, it could be found that there is nearly no amplitude difference in the reflected EM waves between the "0" and "1" states at around 3.5 GHz, while the corresponding phase difference is approximately 180 degrees. As a result, this meta-atom can be effectively utilized in the design of a phase-based digital RIS.

	\section{RIS-aided Passive Human Sensing}
	In this section, we discuss three typical passive human sensing applications with the aid of RISs: imaging and segmentation, localization and tracking, and human activity recognition (HAR). Additionally, we provide an overview of existing algorithms for passive human sensing enhanced by RISs in Table \ref{t1}.
	
	\begin{table*}[h]
		\renewcommand\arraystretch{1.2}
		\centering
		\caption{RIS-aided Passive Human Sensing}
		\begin{threeparttable}
			\begin{tabular}{llllllll}
				\hline
				\begin{tabular}[c]{@{}l@{}}Ref. \\ year\end{tabular}                                     & Applications                   & RIS coding    & Signal frequency  & Sensing algorithm                                                                                            & Multi-person & Real-time & Performance summary                                      \\
				\hline
				\begin{tabular}[c]{@{}l@{}}\cite{li2019machine}\\ 2019\end{tabular}     & Imaging                  & SDC & 3 GHz                        & \begin{tabular}[c]{@{}l@{}}PCA\end{tabular} & No           & Yes       & \begin{tabular}[c]{@{}l@{}} Measured results: $\sim$5 dB SNR of\\ the reconstructed images with 300\\ measurements.  \end{tabular}       \\
				\hline
				\begin{tabular}[c]{@{}l@{}}\cite{li2019intelligent}    \\2019     \end{tabular}                                        & Imaging \&HAR                            & SDC                                                         & 2.4 GHz         & DNN                                                                                                        & Yes           & Yes       & \begin{tabular}[c]{@{}l@{}} Measured results: an accuracy of\\ 0.94 for recognizing ten hand \\signs with WiFi signals, and 0.75\\ SSIM for imaging with 30 coding\\ patterns. \\ \end{tabular}  \\
				\hline 
				\begin{tabular}[c]{@{}l@{}}\cite{li2020intelligent}\\ 2020\end{tabular} & Imaging                  & SDC & 2.4 GHz                  & ANNs                                                                                   & No           & Yes       & \begin{tabular}[c]{@{}l@{}}Measured results: $\sim$0.9 SSIM for\\ imaging with 15 coding patterns. \end{tabular}   \\
				\hline
				\begin{tabular}[c]{@{}l@{}}\cite{hu2022metasketch}\\ 2022\end{tabular}  & Segmentation & SDC & 3.2 GHz                   & MLPs                                                                     & No           & N/A       & \begin{tabular}[c]{@{}l@{}}Measured results: an average error \\rate of \textless{}=1\% for image segmentation.\end{tabular} \\
				\hline
				\begin{tabular}[c]{@{}l@{}}\cite{hu2021metasensing}  \\2021     \end{tabular}                                         & 3D Localization                & SDC                                                         & N/A              & DRL                                                                                                          & No           & N/A       & \begin{tabular}[c]{@{}l@{}}Simulated results: CE loss of 0.9\\ with 16 space grids in 2D space;\\ CE loss of 1.5 with 32 space grids\\ in 3D space.  \end{tabular}                  \\
				\hline
				\begin{tabular}[c]{@{}l@{}}\cite{shao2022target}\\ 2022\end{tabular}    & Localization                   & \begin{tabular}[c]{@{}l@{}}SDC with \\ sensors\end{tabular} & N/A              & MUSIC                                                                                                        & Yes          & N/A       & \begin{tabular}[c]{@{}l@{}}Simulated results: RMSE of $10^{-2}$\\ degree with 15 dBm transmit power. \end{tabular}        \\
				
				\hline
				\begin{tabular}[c]{@{}l@{}}\cite{zhang2022metaradar}  \\2022     \end{tabular}                                        & Localization                   & SDC                                                         & 3 GHz            & WPSO                                                                                                         & Yes          & N/A       & \begin{tabular}[c]{@{}l@{}} Simulated results: localization proba\\-bility of approximately 0.99 with 20\\ detection cycles. \end{tabular}                                                      \\
				\hline
				\begin{tabular}[c]{@{}l@{}}\cite{hu2020reconfigurable} \\2020     \end{tabular}                                  & \begin{tabular}[c]{@{}l@{}}Static posture\\ recognition  \end{tabular}     & SDC                                                         & 3.2 GHz          & FCAO \& DNN                                                                                                  & No           & N/A       & \begin{tabular}[c]{@{}l@{}}Measured results: an accuracy of\\ 0.97  in recognizing four human\\ postures. \end{tabular}        \\
				\hline
				\begin{tabular}[c]{@{}l@{}}\cite{lan2021metasense}    \\2021     \end{tabular}                                        & \begin{tabular}[c]{@{}l@{}}Writing motion\\ recognition \end{tabular}                            & SDC                                                         & 19.4 GHz         & AEMML                                                                                                        & No           & N/A       & \begin{tabular}[c]{@{}l@{}}Measured results: an accuracy of \\over 0.93 for recognizing ten writing\\ motions. \end{tabular}  \\
				\hline  
				\begin{tabular}[c]{@{}l@{}}\cite{wang2022intelligent}   \\2022     \end{tabular}                                      & Activity recognition                            & SDC                                                         & 2.4 GHz          & DNN                                                                                                          & No           & Yes       &  \begin{tabular}[c]{@{}l@{}}Measured results: Imaging with appro\\-ximately 50\% reconstructed images\\ having a normalized PSNR of 75. The\\ normalized PSNR value ranges from 1\\ to 100. \end{tabular}                                                     \\
				\hline

			\end{tabular}
			\begin{tablenotes}
				\footnotesize
				\item SSIM: structure similarity index metric; RMSE: root mean square error; CE: cross-entropy; PSNR: peak signal-to-noise ratio.
			\end{tablenotes}
		\end{threeparttable}
		\label{t1}
	\end{table*}
	
	\subsection{Imaging and Segmentation} 
	RISs are capable of compensating for the limited spatial resolution in current wireless sensing systems, and have made significant advancements in human target imaging. Compared with the commonly used imaging schemes (i.e., real-aperture, synthetic aperture, and coding aperture imaging) that often face challenges in balancing hardware complexity and sensing algorithms, RIS-empowered imaging techniques provide a more efficient and simplified approach to achieve accurate target imaging. For instance, L. Li et al. \cite{li2019machine} proposed a machine-learning reprogrammable imager with RISs. In this work, a linear relationship between the measured reflected signal \textbf{y} and the scene \textbf{x} to be imaged was established, i.e., \textbf{y}=\textbf{H}\textbf{x}. Then, principal component analysis (PCA) was employed to extract
    principle scene components serving as illumination patterns \textbf{H}. On top of that, a deep-learning (DL) based imaging algorithm was developed for data-driven intelligent imaging \cite{li2020intelligent}. The RIS configuration was treated as trainable physical weights and trained with the digital weights of an artificial neural network (ANN) alternatively. Meanwhile, a RIS-empowered imaging and scene segmentation approach was presented \cite{hu2022metasketch}. By optimizing coding sequence of the RIS, the measurements of the customized radio channel \textbf{H} could be utilized to estimate the reflection coefficients of objects. Subsequently, a symmetric multilayer perceptron (MLP) model was proposed to segment different objects using the estimated reflection coefficients.  
	
	RIS-aided human imaging techniques can significantly promote the development of passive human sensing. In other words, the reconstructed images could intuitively capture the physical characteristic of the person, and further infer his body posture, activity and even intention. Meanwhile, the presence of multiple persons could be detected with RIS-aided imaging, which is expected to promote multi-person passive sensing. Furthermore, scene imaging enables a comprehensive understanding of the environment so as to generate desirable signal beams for further fine-grained sensing. 
	
	\subsection{Localization and Tracking}
	
	In RF-based human localization, the target location is generally determined by forming triangles from known points to the target person. In this circumstance, multiple transmitting/receiving antennas are required to improve the localization accuracy, which inevitably increases the system's hardware complexity and deployment difficulty. By contrast, RISs with dense arrays of unit cells could increase the number of antennas to promote localization in a more efficient manner. In addition to the direct link between the target and the Tx, the RISs could provide extra LoS reflected links to perceive the target from other orientations, thus potentially enhancing localization performance. The benefits above have spurred active research on designing various RIS-aided human localization systems. For instance, a semi-passive RIS-empowered localization system, where the RIS is equipped with dedicated sensors to receive echoes for target localization, has been proposed \cite{shao2022target}. The multiple signal classification (MUSIC) algorithm was applied to estimate the direction-of-angle (DOA) of the person with the received echo signals.
	
	To provide desirable signal propagation properties and achieve high sensing accuracy, it is critical to optimize the coding patterns of the RIS. Conventional RIS-assisted localization algorithms generally treat coding sequence optimization and human imaging as separate issues, and the interaction between them has not been considered. To address this limitation, a deep reinforcement learning (DRL) algorithm was proposed to jointly compute the optimal coding patterns and the mapping of the received signals \cite{hu2021metasensing}. Through the joint optimization of hardware and software, the proposed approach was able to sense the presence and locations of 3D objects with a remarkable performance. Meanwhile, H. Zhang et al. \cite{zhang2022metaradar} proposed a waveform and phase shift optimization (WPSO) algorithm for optimizing the radar waveforms and the RIS. 

	\subsection{Human Activity Recognition}
	
	RF-based HAR aims to automatically recognize different human activities by analyzing the impacts of body movements on wireless signal propagation. Conventional RF sensing techniques need to passively adapt to and are generally constrained by the radio environment, which limits the diversity of wireless channels for carrying abundant information about human activities. Furthermore, due to the unpredictable nature of wireless environments, the feasibility and accuracy of activity recognition are greatly affected by the undesirable multi-path fading. To deal with this challenge, RIS-aided HAR techniques have been increasingly investigated. In HAR tasks, the RIS is able to continuously customize wireless propagation properties and generate a high dimension of independent channel measurements to enhance activity recognition performance.
	
	J. Hu et al. \cite{hu2020reconfigurable} proposed a RIS-assisted RF sensing system for recognizing daily human activities. A frame configuration alternating optimization (FCAO) algorithm was first proposed for finding a set of optimal coding sequences of the RIS. Subsequently, a supervised deep neural network (DNN) was proposed to extract features from the received channel information and inference human activities. Similarly, a dynamic metasurface antenna (DMA) was presented for writing motion recognition \cite{lan2021metasense}. The concept ''normalized entropy" was proposed to assess the sensing quality of different DMA pattern configurations and select the optimal patterns. Furthermore, a quality-aware auxiliary-assisted ensemble multimask learning (AEMML) algorithm was presented to dynamically aggregate the heterogeneous DMA measurements for better sensing performance. Additionally, to explore the time-varying characteristics of human behaviors, a RIS-assisted ``camera" \cite{wang2022intelligent} was proposed to produce microwave videos of moving persons in real-time. Experimental results showed that the EM camera successfully recognized human activities behind a 60 cm-thickness reinforced concrete wall with desirable performance in terms of image quality and frame rate.
	\section{Over-the-air Experiments and Results}
	\label{Experiments}
	While significant advancements have been achieved in intelligent RIS-aided human macroscopic activity sensing, there has been limited research on micro-motion (e.g., vital signs) recognition. In this section, we study multi-person respiration monitoring enhanced by SDC and STC RISs, and present some preliminary experimental results.
	
	\subsection{Experimental Setup}
\begin{figure*}[!t]
		\centering
		\includegraphics[width=1.3\columnwidth]{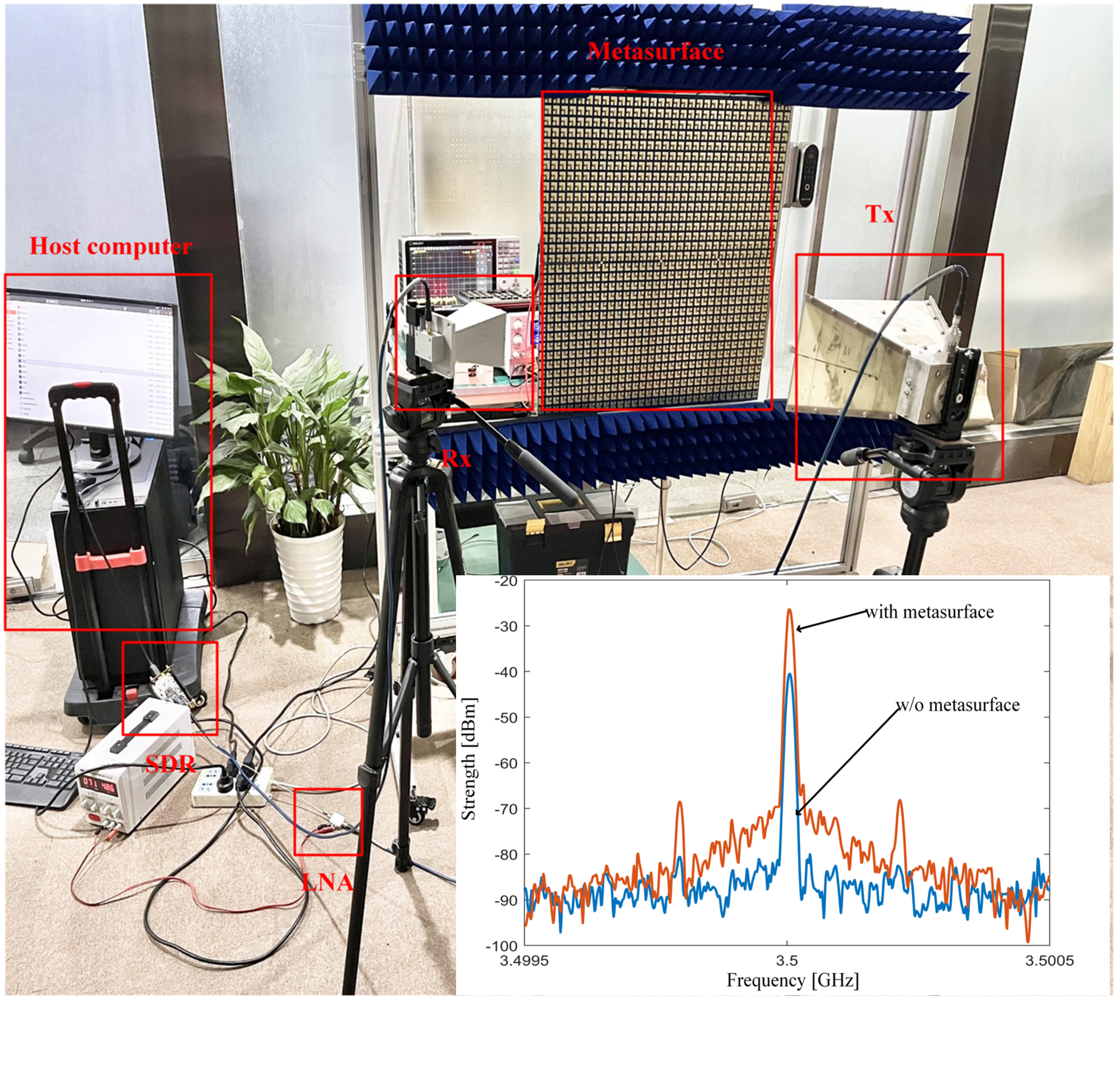}
		\caption{Experimental setup of the RIS-enhanced RF sensing system, which consists of a transceiver module and a RIS module. When the RIS is deployed in this system, the signal gain at the receiving end has been enhanced by over 10 dBm. }
		\label{fig1}
	\end{figure*}

	As shown in Fig. \ref{fig1}, the RIS-enhanced RF sensing system consists of a transceiver module and a RIS module. In the transceiver module, a software-defined radio (SDR) device is adopted and connected with two antennas for signal transmission and reception. Two low-noise amplifiers (LNAs) are connected with the Tx and Rx antennas to amplify the transmitted and received signals of the SDR, respectively. Meanwhile, a host computer is integrated into the system to control the SDR based on the GNU packet and also serves as a data processor for vital signs sensing.
	
	To achieve contactless physiological signal sensing, a 1 kHz baseband tone signal is generated using the GNU packet, and transmitted on the 3.5 GHz carrier frequency through the SDR. The transmitted tone signal can be represented as $Ae^{j(2\pi f_ct)}$, where $A$ is the signal amplitude, and $f_c$ is the carrier frequency, i.e., (3.5+1e-6) GHz. Subsequently, the RIS manipulates the illuminated signals, redirecting them towards the human body. The movements of the human thorax during respiration modify the magnitude of the transmitted signal, leading to amplitude modulation. In other words, the signal received by the SDR could be expressed as $m(t)Ae^{j(2\pi f_ct)}$, where $m(t)$ represents the respiration signal that modulates the transmitted tone signal. To extract $m(t)$ from the demodulated baseband signal, the signal amplitude is first reconstructed with the in-phase (I) and quadrature (Q) components of the received signal and then, the envelope of the amplitude could be a desirable approximation of $m(t)$.
	Subsequently, as the normal respiration rate of a person is 12 to 20 beats per minute, the envelope data is downsampled to 200 Hz. Then, a low pass filter is employed to filtered out the high-frequency noise. Finally, a peak detection algorithm is proposed to estimate human respiration rate.
	
	\subsection{Experimental Results and Analysis}
	
\begin{figure}[!t]
	\centering
	\includegraphics[width=\columnwidth]{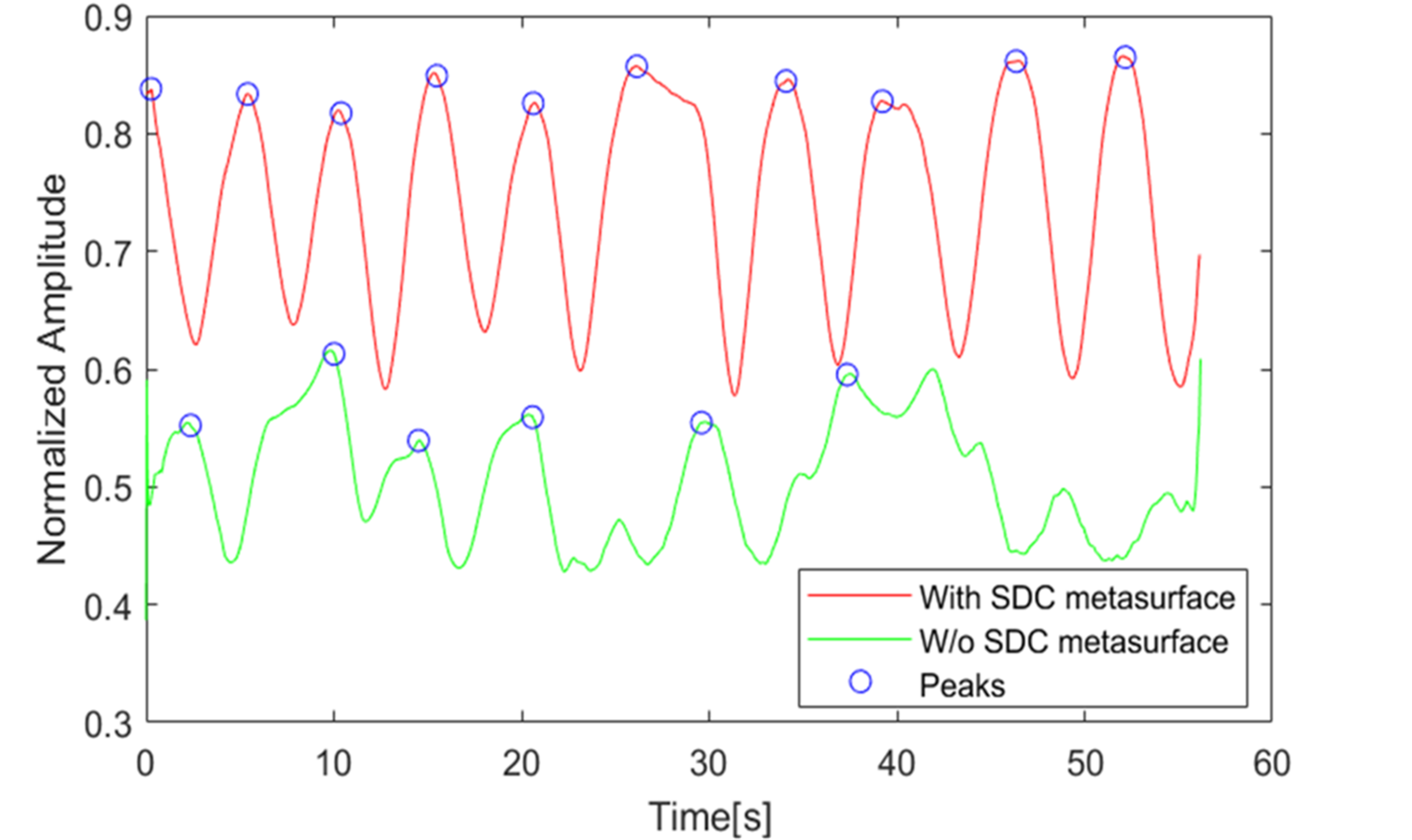}
	\caption{The extracted respiration signals with and w/o the SDC RIS. Compared with the reflected signal when the SDC RIS is not integrated, the signal from the RIS-enhanced system has more distinct periodic properties.}
	\label{fig7}
    \end{figure}

	First, we utilize an SDC RIS to manipulate EM waves and perform single-person respiration sensing. With the experimental deployment in Fig. \ref{fig1}, the signal gain at the receiving end has been enhanced by over 10 dBm, indicating the SDC RIS is capable of significantly improving the signal quality and potentially enhancing sensing performance. Furthermore, when there is a person in the reflection area, the extracted signal $m(t)_{D1}$ after low-pass filtering is shown by the red line in Fig. \ref{fig7}. Compared with the extracted signal $m(t)_{D2}$ (the green line in Fig. \ref{fig7}) when the SDC RIS is not integrated, $m(t)_{D1}$ has more distinct periodic properties, leading to more accurate estimation of the respiration rate. The experimental results indicate that by redirecting signal beams towards the human body with the SDC RIS, other objects in the open space will reflect less signal, thus reducing the interference of the human echo signals at the physical layer and improving the SNR of sensing signals.

	On the other hand, when there are multiple persons in the reflection area of the SDC RIS-assisted respiration monitoring system, the received echoes of these persons become intertwined in the time, frequency and spatial domains. Consequently, it becomes challenging to disentangle these mixed echoes at the receiving end and simultaneously monitor the respiration of multiple individuals using the SDC RIS. To achieve multi-person respiration monitoring, an STC RIS could be incorporated into the RF sensing system instead of the SDC RIS. As the STC RIS is able to modulate the incident wave into a set of harmonic components with different reflection directions, the harmonic waves could redirect the persons at different locations, respectively. Specifically, we assign the harmonic components with different frequencies to different human targets, and align the main beam of each harmonic to the corresponding individual. Then, the reflected echoes from different persons could be separated effectively in the frequency domain. 
	
	\begin{figure*}[!t]
	\centering
	\includegraphics[width=1.8\columnwidth]{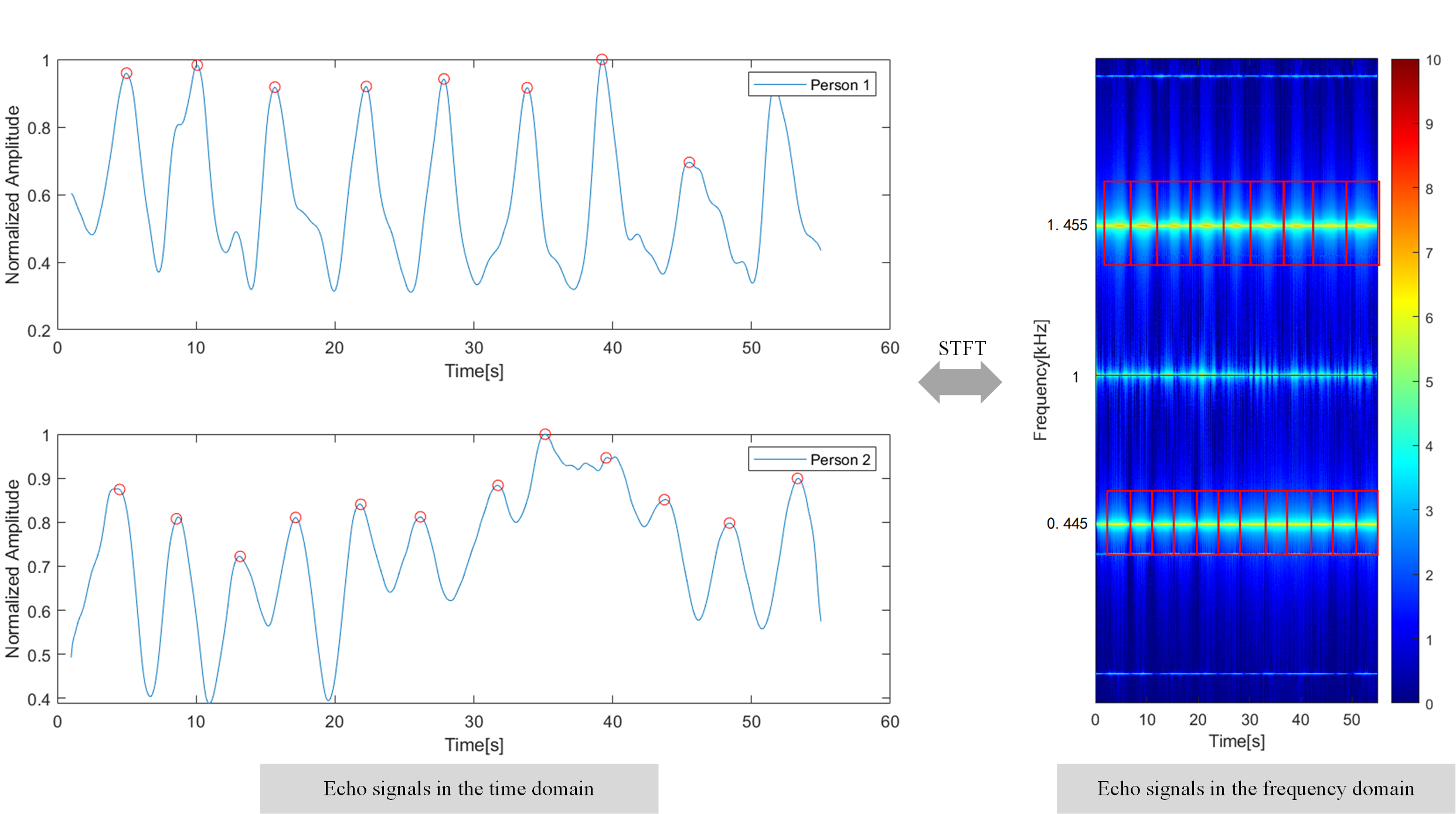}
	\caption{The extracted respiration signals of two persons in the time and the frequency domains. }
	\label{fig4}
    \end{figure*}

	To demonstrate the feasibility and effectiveness of STC RIS-assisted multi-person respiration estimation, an STC RIS with an operating frequency of approximately 3.5 GHz is integrated into the RF sensing system. In the experiment, two persons were positioned respectively at 45 degrees on both sides of the RIS normal. By using a well-designed STC sequences, the reflected beams at the +1st ($f_c$ + 455 Hz) and -1st ($f_c$ - 455 Hz) harmonic frequencies are transmitted along the two directions and illuminate the two persons for sensing. By using only one Rx, the reflected echo signals $m(t)_{T1}$ and $m(t)_{T2}$ from the two persons are received and then separated in the frequency domain by using band-pass filters. The envelope of $m(t)_{T1}$ and $m(t)_{T2}$ are shown in Fig. \ref{fig4}. Meanwhile, the short-time Fourier Transform (STFT) is conducted on $m(t)_{T1}$ and $m(t)_{T2}$ to obtain the time-frequency map, which could also indicate the periodic movements of the chest due to breathing. Experimental results show that the extracted signals have the same periodicity as human breathing, demonstrating the potential of using STC RISs for multi-person respiration monitoring. 

\section{Challenges and Opportunities}
There are a number of open problems in the field of passive human sensing with the aid of RISs. In this section, we briefly discuss several research challenges and future opportunities.

\subsection{Multi-person Activity Sensing}
State-of-the-art RF-based human sensing techniques have achieved excellent performance in single-person activity sensing. However, when it comes to sensing multiple persons, the reflected signals from different persons are generally coupled together, degrading the performance of these approaches. To address this limitation, RISs could be used to separate the mixed reflected echoes. 

For instance, the SDC RIS could manipulate EM beams through time-division multiplexing so that the signal can irradiate different persons in turn to achieve multi-person sensing. However, such a time-division multiplexing sensing scheme exhibits low efficiency, which limits real-time performance. Additionally, the potential of STC RISs for multi-person respiration monitoring has been preliminary verified in Section \ref{Experiments}. These STC RISs possess echo separation capabilities at the physical layer level, which could be further applied for multi-person sensing. Furthermore, in the field of active RF sensing, RISs have been successfully employed to localize multiple mobile users 
, which could provide some inspiration for designing passive multi-person sensing algorithms.

\subsection{Vital Signs Monitoring of Moving Persons}
Current passive RF sensing systems predominantly focus on vital signs monitoring of static persons. When these approaches are utilized for vital signs sensing of moving persons, the performance may degrade significantly. It is because the signals reflected by the human torso and limbs are mixed with physiological signals at the receiving end, and the physiological signals may even be overwhelmed by the reflected echoes from macroscopic body movements. In this case, directing the EM signals towards a specific body part could be a potential solution to eliminate the effect of macroscopic human movements \cite{li2019intelligent}.

More specifically, taking the respiration sensing task as an example, since the moving speed of the thorax is determined by both human macro- and micro- movements, it is essential to compensate for the effect of macro-motions on the sensing measurements, and estimate the distance/speed variations of the thorax caused by respiration alone for the sensing purpose. In future research, we will elaborate on the algorithms for micro-movement estimation and realize vital signs sensing of moving persons.

\subsection{Model-driven Human Sensing with RISs}
Model-driven RF-based human sensing is achieved by quantitatively estimating the effects of the human body on wireless signal propagation (e.g., reflection, diffraction and scattering). Compared with the data-driven human sensing technique, the model-driven alternative does not rely on large-scale data and has strong interpretability, which hence is suitable for sensing scenarios that require low latency and high reliability.

However, when a RIS is integrated into the RF system, the amplitude and phase of channel indicators are simultaneously impacted by the RIS and human movements. In this circumstance, it is vital to decouple the influence of the two factors and individually extract the channel variations caused by human bodies for passive human sensing. Furthermore, though the changes of channel properties caused by RISs could be theoretically derived, accurately qualifying the combined influence of multiple meta-atoms in a real RF system poses practical challenges. Hence it is nontrivial to focus more on the RIS design and the RF system deployment.

\subsection{Joint Optimization of RIS Configuration and Sensing Algorithm}

Jointly optimizing the variables in both the physical and digital layers enables the sensing pipeline to be task-aware and could select more effective information relevant to the task from the received RF signals. To this end, the DL technique is expected to create an end-to-end pipeline for jointly optimizing the physical and digital variables. However, the digitalization characteristic of the meta-atoms introduces discontinuous physical weights, leading to an incompatibility with the existing error backpropagation algorithms. Therefore, it is crucial to develop tailored DL models for RIS configuration optimization. 

Additionally, exploiting information from previous measurements can be a promising approach to acquire optimal system weights and extract more efficient information from the received echoes for sensing tasks. This approach draws inspiration from the recurrent visual attention of humans. In other words, the update of RIS configuration could be impacted based on both the current information and the previous sensing results provided by the digital layer, thereby enabling adaptive customization of more desirable wireless environment for the current sensing task. 

\subsection{Privacy Security of Passive Human Sensing}
Due to the broadcast nature of wireless signal propagation, it is likely that the privacy leakage cannot be prevented in passive human sensing. In this case, the sensitive information such as the human locations, activities and vital signs could be exposed to the air interface, where the malicious users can intercept such sensitive information and further violate user privacy. 

To deal with this issue, an integrated human sensing and security design can be performed, where the RIS-aided sensing and the privacy security can mutually benefit by sharing spectrum, power and hardware resources, etc. For example, as per the illegal user sensing, the RIS can be equipped with active reflection elements, i.e., active RIS, to authenticate the user identity via the radio-frequency fingerprint (RFF) or the physical unclonable function (PUF), and then control the electromagnetic waves to confuse the unauthorized sensing. The challenges mainly include the unified framework design of integrated human sensing and security, especially for the open wireless environment, and the efficient acquisition of RFF or PUF with a low overhead.

\section{Conclusion}
A comprehensive review of RF-based human sensing with the assistance of RISs was presented here. We outlined the background of intelligent metasurfaces and gave an in-depth analysis on the sensing characteristics of the SDC and STC RISs. Subsequently, we categorized the existing RIS-aided passive human sensing systems into three typical applications, followed by a detailed discussion on their deployments and implementations for the sensing purpose. Furthermore, RF-based human respiration monitoring assisted by SDC and STC RISs was investigated. The experimental setup was introduced briefly, and some experimental results were then provided to demonstrate the potential of using STC RISs for multi-person micro-motion sensing. Finally, we presented four key challenges and open research problems to facilitate a transition of the techniques into real-world applications.

\section*{Acknowledgements}
This work was supported in part by the Postdoctoral Innovation Talents Support Program under Grant BX20230066, in part by the Jiangsu Planned Projects for Postdoctoral Research Fund under Grant 2023ZB318, in part by the National Natural Science Foundation of China under Grant 62371131, 62288101, 62371131, and 62371132, in part by the Natural Science Foundation of Jiangsu Province under Grant BK20230820, and in part by Fundamental Research Funds for the Central Universities under Grant 2242023K5002, and in part by the China Postdoctoral Science Foundation under Grant 2021M700761 and 2022T150112.

\bibliographystyle{IEEEtran}

\bibliography{refs}

\vfill

\end{document}